\documentclass[conference, final]{IEEEtran}

\usepackage[utf8]{inputenc}

\usepackage[acronym,shortcuts]{glossaries}
\usepackage{bm}
\usepackage{amsmath, amsthm, amssymb}
\usepackage{xcolor}
\usepackage{soul}
\usepackage{enumitem}
\usepackage[caption=false]{subfig}

\usepackage{pgfplots}
\pgfplotsset{compat=newest}
\pgfplotsset{plot coordinates/math parser=false}
\usetikzlibrary{plotmarks}
\usetikzlibrary{arrows}
\usetikzlibrary{decorations}
\usepackage{grffile}
\pgfplotsset{plot coordinates/math parser=false}

\newlength\fwidth
\newlength\fheight
\newacronym{uwan}{UWAN}{underwater acoustic network}
\newacronym{id}{ID}{identification number}
\newacronym{uwac}{UWAC}{underwater acoustic channel}
\newacronym{fa}{FA}{false alarm}
\newacronym{ld}{LD}{local decision}
\newacronym{md}{MD}{missed detection}
\newacronym{mse}{MSE}{mean square error}
\newacronym{pdf}{PDF}{probability density function}
\newacronym{cdf}{CDF}{cumulative distribution function}
\newacronym{ml}{ML}{machine learning}
\newacronym{nn}{NN}{neural network}
\newacronym{ae}{AE}{autoencoder NN}
\newacronym{rms}{RMS}{root mean square}
\newacronym{ssp}{SSP}{sound speed profiles}
\newacronym{gmm}{GMM}{Gaussian mixture model}
\newacronym{lrt}{LRT}{likelihood ratio test}
\newacronym{glrt}{GLRT}{generalized likelihood ratio test}
\newacronym{iid}{i.i.d.}{independent and identically distributed}
\newacronym{pls}{PLS}{physical-layer security}
\newacronym{kde}{KDE}{kernel density estimation}
\newacronym{relu}{ReLu}{rectified linear unit}
\newacronym{svm}{SVM}{support vector machine}
\newacronym{crc}{CRC}{cyclic redundancy check}
\newacronym{arq}{ARQ}{automatic repeat request}

\title{Machine Learning-Based Distributed Authentication of UWAN Nodes with Limited Shared Information}
\author{Francesco~Ardizzon$^{1*}$, Roee~Diamant$^2$, Paolo~Casari$^3$, and~Stefano~Tomasin$^1$ \\
\small
$^1$University of Padova and CNIT, Italy, \;\; $^2$University of Haifa, Israel, \;\; $^3$University of Trento and CNIT, Italy \\ *Corresponding author, email:francesco.ardizzon@phd.unipd.it
\thanks{This work was sponsored in part by the NATO Science for Peace and Security Programme under grant no.~G5884 (SAFE-UComm), and by MIUR (Italian Ministry of Education) under the initiative {\it Departments of Excellence} (Law 232/2016).}
 
}
\date{}
\IEEEoverridecommandlockouts
\begin{document}
\maketitle
 
\begin{abstract}
We propose a technique to authenticate received packets in underwater acoustic networks based on the physical-layer features of the \ac{uwac}. Several sensors a) locally estimate features (e.g., the number of taps or the delay spread) of the \ac{uwac} over which the packet is received, b) obtain a {\em compressed feature representation} through a \ac{nn}, and c) transmit their representations to a central sink node that, using a \ac{nn}, decides whether the packet has been transmitted by the legitimate node or by an impersonating attacker. Although the purpose of the system is to make a binary decision as to whether a packet is authentic or not, we show the importance of having a rich set of compressed features, while still taking into account transmission rate limits among the nodes. We consider both global training, where all \acp{nn} are trained together, and local training, where each \ac{nn} is trained individually. For the latter scenario, several alternatives for the \ac{nn} structure and loss function were used for training. 
\end{abstract}
\glsresetall

\section{Introduction}\label{sec:intro}

\Ac{pls} is an emerging approach that can potentially withstand quantum computing attacks (as it is not founded on computationally hard problems) and is particularly suitable in networks with communication or computing constraints. We focus on \ac{pls} for the authentication of \ac{uwan} packets, wherein a receiving device must decide whether a received packet has been transmitted by the claimed transmitter, or by an {\em impersonating} transmitter aiming at sending fake and potentially dangerous packets to the receiving device. 

Authentication  has been studied in the literature from its theoretical foundation as a hypothesis testing problem \cite{850674}, to its practical implementation \cite{6204019} and performance limits \cite{7010914,7270404}. In \cite{4697498}, the \ac{uwac} is used to generate secret keys, then for authentication (for a comparison with key-less authentication, see also~\cite{8396873}). In \cite{9676618}, an authentication scheme using the maximum time-reversal resonating strength of \ac{uwac} is proposed. In \cite{cooperative19} several trusted nodes independently help a sink node in the authentication process, by computing one belief value each (based on estimated channel parameters). Beliefs are then fused by a sink node to make a decision. Both the statistical parameters and the fusion are obtained by assuming a particular statistical model. However, such a model may not always be accurate, considering also the large variety of scenarios for \ac{uwan} with diverse channel features \cite{4752682}. In this respect, \ac{ml} approaches that {\em learn} the channel statistics constitute a promising solution. The authentication scheme of  \cite{8501958}  provides {\em a single receiver} that exploits reinforcement learning to choose the authentication parameter without being aware of the network and spoofing model. Still, a single receiver may not provide an accurate authentication process. Therefore, solutions based on multiple devices have been explored. In \cite{8823047} a general framework for a synergic trust model based on a \ac{svm} is proposed. However, this scheme is complex and is not tailored to detect spoofing attacks. 

In this paper, we build upon~\cite{comcas21}, and consider a set of trusted nodes that report information to a central node making the final authentication decision. In~\cite{comcas21}, the trusted node only reported a single value for each packet, which represents a local (soft) decision: however, this choice limits the ability of the central node to effectively fuse the information, e.g., taking into account the correlation of the features computed by different trusted nodes. Indeed, we advocate that passing more values (still compressed with respect to all the observed features) yields a more accurate decision. We resort to \ac{ml}, where both trusted nodes use \acp{nn} to compress the data (locally) and the central node uses a \ac{nn} to centrally fuse the information. Although the purpose of the system is to take a binary decision (whether a packet is authentic or not) we show the importance of having a rich set of compressed features, still taking into account the transmission rate limits among the sensors and the central nodes. We consider both a global training where all \acp{nn} are trained together, and localized training, where the \ac{nn} at each sensor is first trained and then the central \ac{nn} is trained with fixed local \acp{nn}.

\section{System Model}\label{sec:dataModeling}

\begin{figure}
    \centering
    \resizebox{.875\columnwidth}{!}{%
    \input{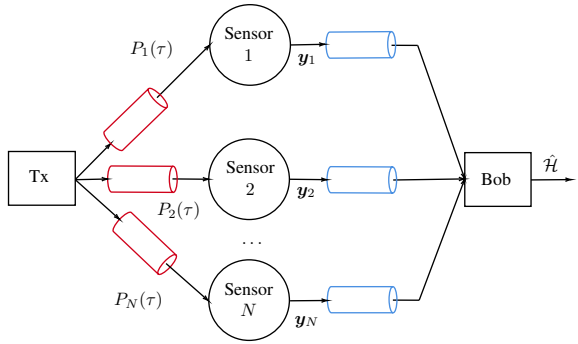}
     }
    \caption{Signal model of the proposed \ac{uwan} authentication scheme.}
    \label{fig:high_level_scheme}
        \vspace{-2mm}

\end{figure}

We consider the scenario of Fig.~\ref{fig:high_level_scheme}, where $N$ {\em sensors} cooperate with a node (Bob) who decides if the received packet originate from a specific sender (Alice) or not. In turn, the attacker (Eve) transmits packets to the sensors in an attempt to impersonate Alice, i.e., aiming at having Bob accepting her packets as coming from Alice. Communications occur over \acp{uwac}: the transmissions from the sensors to Bob are performed over authenticated channels, i.e., Eve cannot transmit signals over these channels. {We also assume that the sensors and Bob employ proper error detection and correction protocols (e.g., \ac{crc} and \ac{arq}), such that no communication error occurs in the data reception. Moreover, Eve does not modify the transmit signal to specifically break authentication (more sophisticated attacks are left for future study).}
 
\section{Authentication Protocol}\label{sec:authProtocol}

We propose the following authentication protocol upon the reception of a packet by the network: 
\begin{enumerate}
    \item each sensor $n=1, \ldots, N$, estimates the power delay profile $\{P_n(\tau)\}$, i.e., the power of tap with delay $\tau$, of the channel over which the packet was received,
    \item each sensor $n=1, \ldots, N$, extracts the feature vector $\bm{x}_n \in \mathbb{R}^K$ from  $\{P_n(\tau)\}$,
    \item  each sensor $n=1, \ldots, N$, processes $\bm{x}_n$ to obtain the (compressed) vector $\bm{y}_n = f_n(\bm{x}_n)  \in \mathbb{R}^M$ and $M<K$, 
    \item Bob collects the $N$ local outputs $\bm{y}_n$, $n=1, \ldots, N$, computes $z = g([\bm{y}_1,\ldots,\bm{y}_N])$, and 
    \item Bob verifies the authenticity of the packet through the following test $\hat{\mathcal{H}} = 1$ if $z \geq \lambda$, and $\hat{\mathcal{H}} = 0$ if $z < \lambda$.
\end{enumerate}
The value of the threshold $\lambda$ is chosen to match the target \ac{fa} probability, $p_{\rm FA}$, hence, indicating with $\mathcal H=1$ ($\mathcal H=0$) the case wherein Alice (Eve) is transmitting, the \ac{fa} probability is $p_{\rm FA} = {\mathbb P}[\hat{\mathcal H} =0| {\mathcal H} =1]$.

Now, functions $f_n(\cdot)$, $n=1, \ldots, N$, and $g(\cdot)$ must be designed to obtain a robust authentication process. Since the distribution of the input feature vector is unknown, we resort to \ac{ml} techniques to design them. In particular, they are implemented as \acp{nn} with multiple layers \cite{Goodfellow-et-al-2016}. 

About the features, in \cite{cooperative19,comcas21} and \cite{9598102} their computation is studied in detail, taking into account the stability of their statistics over time and their strong dependency on the transmitter's location~\cite{cooperative19} to effectively distinguish packets transmitted from different positions. The considered features are the number of channel taps, the average tap power, the relative \ac{rms} delay, and the smoothed received power, as defined in \cite{cooperative19,comcas21}.

\paragraph*{Remark} In \cite{comcas21} we considered a similar authentication mechanism, where however each sensor reports a single value to Bob, i.e., $M=1$. Here instead we consider larger values of $M$. Indeed, if the features of the sensors are correlated, it is advantageous to provide a more detailed description to Bob, with a larger number $M$ of reported values.

We consider now a supervised training with a two-sided dataset, i.e., the training is based on packets coming from both Alice and Eve, and the true label (packet source) is available during training. {Note that our approach requires having prior data from Alice and Eve: such data may be available, e.g., in semi-static contexts, where Alice is at a fixed position and its dataset can be well characterized. For Eve, worst-case scenarios (e.g., when Eve is located very close to Alice) can be considered to collect data.} For training, we consider two possible scenarios: a) local training and b) global training. With local training, each \ac{nn} implementing function $f_n(\cdot)$ is trained separately. Afterwards,, function $g(\cdot)$ is also trained; this does not require communication between sensors and Bob during  training. In the global training scenario, instead, all \acp{nn} are trained together as a single large \ac{nn} including functions $f_n(\cdot)$, $n=1, \ldots, N$, and $g(\cdot)$; thus, the sensors must communicate with Bob during training.

\subsection{Local Training}\label{sec:localTraining}

In local training, each function $f_n(\cdot)$, $n=1, \ldots, N$, is trained  locally at its sensor. We propose three options for the output that these \acp{nn} report to Bob. 

\paragraph*{\Ac{ae} solution} The first solution is to perform a lossy source coding on the features using as loss function the \ac{mse} on the reconstructed vector $\tilde{\bm{x}}_n$, i.e., ${\mathbb E}[||\tilde{\bm{x}}_n - \bm{x}_n||^2]$. Such compression is obtained by training the local \ac{nn} (operating as the {\em encoder}) in cascade with a second \ac{nn} (operating as the {\em decoder}) that provides $\tilde{\bm{x}}_n$ from $\bm{y}_n$, in what is known as an \ac{ae} (see \cite{Goodfellow-et-al-2016} for more details). Note that this solution is not tailored to the hypothesis testing problem, but it aims at providing the best representation of the observed features to Bob (in terms of \ac{mse}): in fact, the training of the \ac{ae} is {\em unsupervised}, i.e., we do not exploit the true label of the feature.

\begin{figure}
    \centering
    \resizebox{.85\columnwidth}{!}{%
    \input{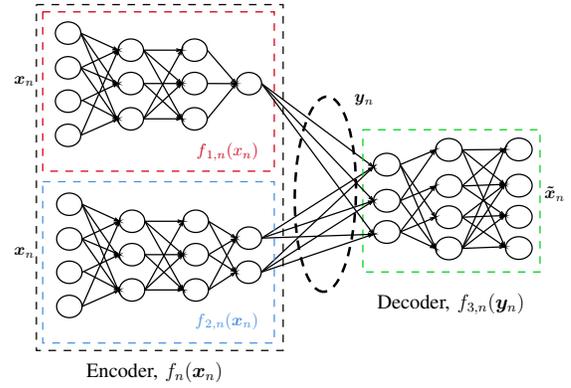}}
    \caption{Example of architecture for the modified \ac{ae} with input size $K=4$ and local output size $M=3$.}
    \label{fig:schemeAE}
        \vspace{-2mm}

\end{figure}

\paragraph*{\Ac{ld} solution} The second solution is provided by \cite{comcas21}, where $M=1$ and the local \ac{nn} is trained to provide the best authentication test at the local level. In this case, the loss function is the \ac{mse} with respect to the true label, i.e., ${\mathbb E}[||y_n - {\mathcal H}||^2]$. Although in this case the compression is targeted to the hypothesis testing problem, the information passed to Bob may not allow for an effective exploitation of the correlation among the features observed at the various sensors.

\paragraph*{Combined \ac{ld} and \ac{ae} (CLDAE) solution} To reap the benefits of both local approaches (\ac{ld} and \ac{ae}) we propose here  a modified version of the \ac{ae}, shown in Fig.~\ref{fig:schemeAE} (for the case $K=4$ and $M=3$). In particular, we split the encoder \ac{nn} $f_n(\bm{x}_n)$ into two \acp{nn} implementing functions $f_{1,n}(\bm{x}_n)$ and $f_{2,n}(\bm{x}_n)$, both having the feature vector $\bm{x}_n$ as input. The first function $f_{1,n}(\bm{x}_n)$ has a single output and implements the local decision as in  \cite{comcas21}. The second function $f_{2,n}(\bm{x}_n)$ has  $M-1$ outputs and the $M$ outputs of both $f_{1,n}(\bm{x}_n)$ and $f_{2,n}(\bm{x}_n)$ are seen as the output of the encoder part of an \ac{ae} providing the reconstructed vector $\tilde{\bm{x}}_n$. Thus, for training we add in cascade a third \ac{nn} $f_{3,n}(\bm{y}_n)$ operating as decoder and train both $f_{n}(\bm{x}_n)$ and $f_{3,n}(\bm{y}_n)$ to minimize the \ac{mse} with respect to the input feature vector, i.e., ${\mathbb E}[||\tilde{\bm{x}}_n - \bm{x}_n||^2]$. 

For all three solutions, function $g(\bm{y}_n)$ is implemented as a \ac{nn}, whose loss function for training is the \ac{mse} with respect to the true label, i.e., ${\mathbb E}[||z - {\mathcal H}||^2]$.

\subsection{Global Training}\label{sec:centrTraining}

With global training, $f_n(\bm{x}_n)$, $n=1, \ldots, N$, and $g(\cdot)$ are jointly trained by using as loss function the \ac{mse} with respect to the true label, i.e., ${\mathbb E}[||z - {\mathcal H}||^2]$. 

\section{Dataset Generation}\label{sec:dataset}
To test the performance of the proposed authentication technique, we performed a sea experiment in January 2022 in Eilat, Israel. This area is characterized by a complex bathymetry and is thus a good environment to test our method, that relies on source separation based on channel features. For network communications, we used 7 Nanomodem-v3 from Newcastle University, UK, commercialized by Succorfish. These low-cost cylindrical modems measure 4~cm $\times$ 6~cm, operate in the 24--32~kHz band, and have a source power level of 168~dB. They can be used to transmit packets of up to 64~bytes, either broadcast or unicast, using 640-bps 16-ary orthogonal modulation signals. To obtain the channels' impulse responses, we used Raspberry~Pi boards to start polling sequences, in which a trusted node transmits a channel-request and in response receives a message from which the magnitude of the channel's taps is obtained with a resolution of 10~$\mu$s.

\begin{figure}
    \centering
    \includegraphics[width=0.84\hsize]{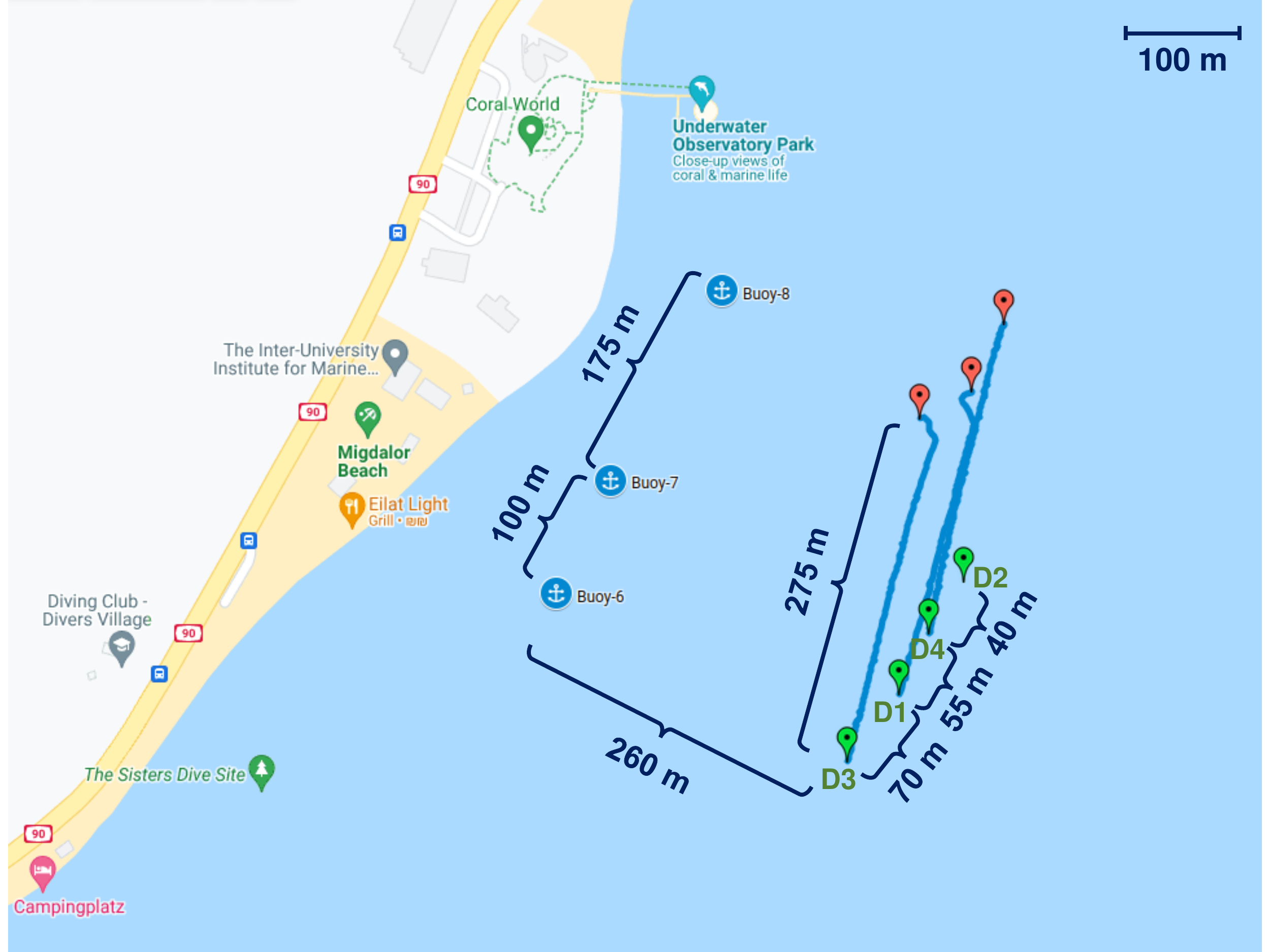}
    \caption{Scheme of the deployment for the sea experiment in Eilat, Israel. {The drifters' trajectories are shown as blue lines, where the red circles mark their position at the end of the experiment. The floaters appear as blue circles}.}
    \label{fig:exptopo}
        \vspace{-2mm}

\end{figure}

As shown in Fig.~\ref{fig:exptopo}, three floaters were chosen as trusted nodes, and communicated with four submerged drifters. The floaters were anchored roughly 150~m apart along a north-south line at a water depth of 40~m. Each floater integrated a Nanomodem deployed at a depth of 5~m. The four drifters were initially deployed at roughly 250~m from the floaters and formed a north-south line with roughly 50~m spacing. The drifters also integrated one Nanomodem each, and were initially deployed at a place where the water column depth is 85~m. From their deployment point, they drifted at roughly 0.25~knots towards north-west, pushed by the water current. While drifting, they maintained a constant depth of 25~m depth. The three trusted nodes initiated polling cycles using a time-division-multiple-access (TDMA) protocol having a 1-minute cycle. 
Here, each floater has a 20~s time window, during which it interrogated each of the four drifters to obtain the current channel impulse response. 


We obtained a dataset including 30~minutes of measurements, which were insufficient to train the \acp{nn}. Therefore, we artificially generated additional features according to the statistics of the collected data: we fitted each data series $x_{n,k}$ using a Gaussian \ac{kde} model, estimating  the \acp{pdf} $p_{x_{n,k}}(x)$. Moreover, we add correlation to the estimated features at different sensors to test the importance of a richer ($M>1$) information transfer from the sensors to Bob.  

To generate correlated features, we adopt the following procedure based on the inverse transform sampling method~\cite{Devroye_1986}:
\begin{enumerate}
    \item we generate a $N \times K$ matrix  of zero-mean correlated Gaussian variables $\tilde{\bm{v}}$, with unitary covariances  $\mbox{COV}(v_{n,k},v_{n',k'}) = 
    1$, if $n=n',k=k'$,  $\mbox{COV}(v_{n,k},v_{n',k'}) = 
    \alpha$, if  $n\neq n',k=k'$, and zero otherwise,
where $\alpha \in [0,1]$ is a parameter to control the covariance among sensors;  
    \item we compute $u_{n,k} = F^{N}_x(\tilde{v}_{n,k})$, where $F^{N}_x(\tilde{v}_{n,k})$ is the \ac{cdf} of a normal distribution, and derive $x_{n,k} = F^{-1}_{x_{n,k}}(u_{n,k})$ via numerical methods.
\end{enumerate}
We consider only the measurements from transmitters 1 and 3: the first is considered to be Alice and the second Eve. Each generated dataset contains $10^5$ measurements per feature. We generated a dataset for each considered value of $\alpha$ and we used 60\% of data set for training, 15\% for validation, and 25\% for testing.

\section{Performance Results}\label{sec:results}

We considered a scenario with $N=3$ sensors, each computing the $K=4$ features described in Section~\ref{sec:dataModeling}. The channel models and the generation of the dataset used for training and performance evaluation have been described in Section~\ref{sec:dataset}.
The neurons of the \acp{nn} use the \ac{relu} as activation function, unless otherwise specified. 

\paragraph*{\Ac{ae} Solution} 
For the encoder \ac{ae} solution, we considered  an input layer with $4$ neurons followed by $2$ layers with $3$ neurons each. The hidden layer has $M = 1$, $2$, or $3$ neurons. The structure of the decoder mirrors the encoder. The output layer has a linear activation function.

\paragraph*{\ac{ld} Solution}
The \ac{ld} solution is used both standalone for the case $M=1$ and as $f_{1,n}(\cdot)$ for the CLDAE solution. We used $4$ layers, with $4$, $3$,  $2$, and $1$ neurons, respectively. The output neuron uses the sigmoid activation function.

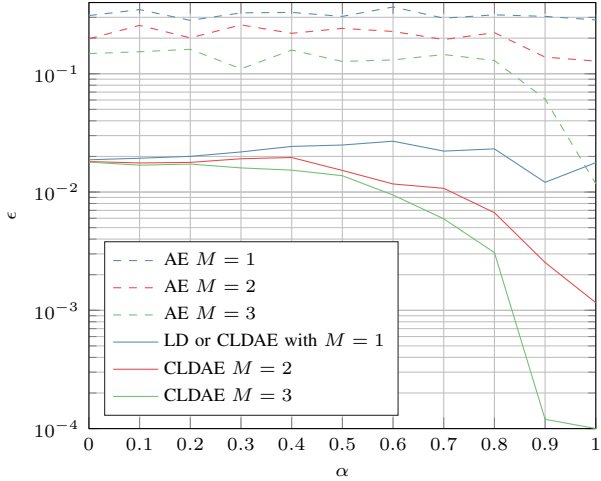
\begin{figure}
    \centering
%
\definecolor{mycolor1}{rgb}{0.34667,0.53600,0.69067}%
\definecolor{mycolor2}{rgb}{0.91529,0.28157,0.28784}%
\definecolor{mycolor3}{rgb}{0.44157,0.74902,0.43216}%
\begin{tikzpicture}

\begin{axis}[%
width=0.951\fwidth,
height=1\fheight,
at={(0\fwidth,0\fheight)},
scale only axis,
xmin=0,
xmax=1,
xlabel style={font=\color{white!15!black}},
xlabel={$\alpha$},
xtick=data,
ymode=log,
ymin=0.0001,
ymax=0.4,
yminorticks=true,
ylabel style={font=\color{white!15!black}},
ylabel={$\epsilon$},
axis background/.style={fill=white},
xmajorgrids,
ymajorgrids,
yminorgrids,
legend style={at={(0.03,0.03)}, anchor=south west, legend cell align=left, align=left, draw=white!15!black},
enlargelimits=false,title style={font=\scriptsize},xlabel style={font=\scriptsize},ylabel style={font=\scriptsize},legend style={font=\scriptsize},ticklabel style={font=\scriptsize}
]
\addplot [color=mycolor1, dashed]
  table[row sep=crcr]{%
0	0.310274279986896\\
0.1	0.348722848399879\\
0.2	0.282577576866646\\
0.3	0.327190808905486\\
0.4	0.330036900471382\\
0.5	0.305444575160394\\
0.6	0.366867499797164\\
0.7	0.295060345276863\\
0.8	0.314853962318593\\
0.9	0.305540418883435\\
1	0.285163156149156\\
};
\addlegendentry{AE $M=1$}

\addplot [color=mycolor2, dashed]
  table[row sep=crcr]{%
0	0.197382271659527\\
0.1	0.256528518167088\\
0.2	0.201063773044086\\
0.3	0.258843974993445\\
0.4	0.219660660171484\\
0.5	0.241885386918732\\
0.6	0.227786543855371\\
0.7	0.194601410866324\\
0.8	0.22253362182044\\
0.9	0.138215603184828\\
1	0.128024719977206\\
};
\addlegendentry{AE $M=2$}

\addplot [color=mycolor3, dashed]
  table[row sep=crcr]{%
0	0.148027717183088\\
0.1	0.153509274128327\\
0.2	0.160946451430925\\
0.3	0.110036972600679\\
0.4	0.158561346152456\\
0.5	0.127129841889446\\
0.6	0.130943793255137\\
0.7	0.145391253385797\\
0.8	0.129275400061432\\
0.9	0.0610998612288652\\
1	0.0117408830728996\\
};
\addlegendentry{AE $M=3$}

\addplot [color=mycolor1]
  table[row sep=crcr]{%
0	0.0187432073578846\\
0.1	0.0193149920372009\\
0.2	0.020027203241049\\
0.3	0.021825200364722\\
0.4	0.0243047361408245\\
0.5	0.0249727065493291\\
0.6	0.0268998643950022\\
0.7	0.0221853764398234\\
0.8	0.023164050034026\\
0.9	0.0120814043946557\\
1	0.0177285059159878\\
};
\addlegendentry{LD or CLDAE with $M=1$}

\addplot [color=mycolor2]
  table[row sep=crcr]{%
0	0.0180909124311872\\
0.1	0.0176174207189428\\
0.2	0.0178124338248103\\
0.3	0.0191199069300388\\
0.4	0.0195813568258004\\
0.5	0.0152250097091344\\
0.6	0.0117015543658637\\
0.7	0.0107491692442651\\
0.8	0.00668867224627423\\
0.9	0.00253775602276274\\
1	0.00116036006048903\\
};
\addlegendentry{CLDAE $M=2$}

\addplot [color=mycolor3]
  table[row sep=crcr]{%
0	0.0179115028058328\\
0.1	0.0168729129344796\\
0.2	0.017236311467513\\
0.3	0.0159956107626921\\
0.4	0.0153174641712439\\
0.5	0.0137389749937968\\
0.6	0.0094316895591422\\
0.7	0.00593730833855432\\
0.8	0.00307734172321861\\
0.9	0.000119832554301702\\
1	0.000100086599316818\\
};
\addlegendentry{CLDAE $M=3$}

\end{axis}
\end{tikzpicture}%
    \vspace{-3mm}
    \caption{Error rate $\epsilon$ for the proposed protocol, for $M=1,2$ and $3$, as a function of $\alpha$ considering the local training scenario.}
    \vspace{-2mm}
    \label{fig:local}
\end{figure}

\paragraph*{CLDAE Solution} For the CLDAE solution, function $f_{2,n}(\cdot)$ is implemented as a \ac{nn} with $3$ layers, having $4$ neurons in the first layer, $3$ in the second, and $M-1$ in the third. The decoder is implemented as in the \ac{ae} solution.

In all the three solutions, the \ac{nn} implementing $g(\cdot)$ was designed with $MN$ neurons in the first layer, $N$ in the second, and a single neuron in the output layer. The output neuron uses the sigmoid activation function.

We characterize the performance of our solutions in terms of the error rate, $\epsilon = {\mathbb P}[\hat{\mathcal H} \neq {\mathcal H}]$, obtained for an attack probability of 0.5, i.e., $\epsilon = 0.5 p_{\rm FA} + 0.5 p_{\rm MD}$, where  $p_{\rm MD} = {\mathbb P}[\hat{\mathcal H} =1| {\mathcal H} =0]$ is the \ac{md} probability and the testing threshold  $\lambda$ has been optimized  to minimize $\epsilon$.

Fig.~\ref{fig:local} shows the error probability of the local solutions in the considered scenario. As $M$ increases, more information on the observations is provided to Bob, who can also exploit the correlation among them, thus reducing $\epsilon$. When comparing the various solutions, we note both LD and CLDAE outperform the \ac{ae} solution, since they are targeted towards the hypothesis testing problem. Moreover, we clearly see that providing more information than only soft local decisions (as would be the case for $M=1$) decreases the error probability, since Bob can exploit the correlation of the observations at the sensors. Lastly, we observe that the error probability decreases as the correlation increases for all solutions. This is due to the fact that statistics at different sensors are different, thus having multiple highly correlated measurements makes it possible to reduce the decision uncertainty.

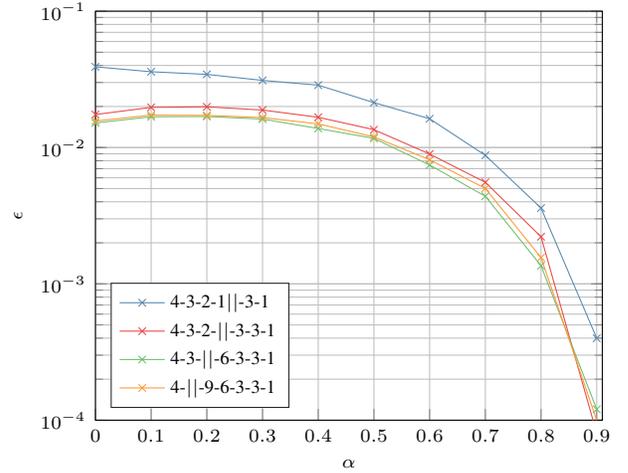
\begin{figure}
    \centering
%
\definecolor{mycolor1}{rgb}{0.34667,0.53600,0.69067}%
\definecolor{mycolor2}{rgb}{0.91529,0.28157,0.28784}%
\definecolor{mycolor3}{rgb}{0.44157,0.74902,0.43216}%
\definecolor{mycolor4}{rgb}{1.00000,0.59843,0.20000}%
\begin{tikzpicture}

\begin{axis}[%
width=0.951\fwidth,
height=0.96\fheight,
at={(0\fwidth,0\fheight)},
scale only axis,
xmin=0,
xmax=0.91,
xlabel style={font=\color{white!15!black}},
xlabel={$\alpha$},
xtick=data,
ymode=log,
ymin=0.0001,
ymax=0.1,
yminorticks=true,
ylabel style={font=\color{white!15!black}},
ylabel={$\epsilon$},
axis background/.style={fill=white},
xmajorgrids,
ymajorgrids,
yminorgrids,
legend style={at={(0.03,0.03)}, anchor=south west, legend cell align=left, align=left, draw=white!15!black},
enlargelimits=false,title style={font=\scriptsize},xlabel style={font=\scriptsize},ylabel style={font=\scriptsize},legend style={font=\scriptsize},ticklabel style={font=\scriptsize}
]
\addplot [color=mycolor1, mark=x, mark options={solid, mycolor1}]
  table[row sep=crcr]{%
0.9	0.000399328778972974\\
0.8	0.00359938630832723\\
0.7	0.00876837043425637\\
0.6	0.0162650284061746\\
0.5	0.0213506190328887\\
0.4	0.0286915862058601\\
0.3	0.0310819619790797\\
0.2	0.0344110826286791\\
0.1	0.0359731459108279\\
0	0.039100566934352\\
};
\addlegendentry{4-3-2-1$|$$|$-3-1}

\addplot [color=mycolor2, mark=x, mark options={solid, mycolor2}]
  table[row sep=crcr]{%
0.9	8.01710412828838e-05\\
0.8	0.00221656304848383\\
0.7	0.00555721670541909\\
0.6	0.00896057228958436\\
0.5	0.0135255727573369\\
0.4	0.0166776582242685\\
0.3	0.0188436331632996\\
0.2	0.0198868743182717\\
0.0999999999999996	0.0197070278850645\\
0	0.0174658003931731\\
};
\addlegendentry{4-3-2-$|$$|$-3-3-1}

\addplot [color=mycolor3, mark=x, mark options={solid, mycolor3}]
  table[row sep=crcr]{%
0.9	0.000120171760399801\\
0.8	0.00135866762558312\\
0.7	0.00439295577480181\\
0.6	0.00744868590949577\\
0.5	0.0116696945931015\\
0.4	0.0137964192265063\\
0.3	0.0161690122800352\\
0.2	0.0168870619888457\\
0.1	0.0168052925188268\\
0	0.0151498291575687\\
};
\addlegendentry{4-3-$|$$|$-6-3-3-1}

\addplot [color=mycolor4, mark=x, mark options={solid, mycolor4}]
  table[row sep=crcr]{%
0.9	9.99169962677681e-05\\
0.8	0.00155968883946211\\
0.7	0.00501644378361976\\
0.6	0.00813124218021942\\
0.5	0.0119759624990634\\
0.4	0.0149110120642105\\
0.3	0.0165929079006612\\
0.2	0.0172109686119093\\
0.0999999999999996	0.0172909700501431\\
0	0.0156473069010246\\
};
\addlegendentry{4-$|$$|$-9-6-3-3-1}

\end{axis}
\end{tikzpicture}%
    \vspace{-3mm}
    \caption{Error rate $\epsilon$ for the proposed protocol, for $M=1,2$ and $3$, as a function of $\alpha$ considering the global training scenario.}
    \vspace{-2mm}
    \label{fig:global}
\end{figure}

\paragraph*{Global training} For global training, we describe the \acp{nn} with notation $a_1-a_2- \ldots - a_{Q_L}||b_1- \ldots-b_{Q_G}$, where $a_p$ indicates the number of neurons of layer $p$ in each local \ac{nn}, while $b_q$ indicates the number of neurons in layer $q$ of Bob's \ac{nn}. The considered configurations for $M=1$, $2$, and $3$, are reported in the legend of Fig.~\ref{fig:global}. Note that in all configurations the total number of neurons is $34$, and the output neurons use the sigmoid activation function.

Fig.~\ref{fig:global} shows the error rate $\epsilon$ as a function of the correlation coefficient $\alpha$ for different \ac{nn} configurations.
Comparing Fig.~\ref{fig:local} and Fig.~\ref{fig:global}, we notice that, even if the latter achieves the best performance, CLDAE with local training achieves almost the same results, even though it operates under more restrictive assumptions: considering, for instance, $\alpha = 0.7$ and $M=2$, by using the \ac{ae} we would always get $\epsilon>10^{-1}$, while with the CLDAE we achieve $\epsilon \approx 1.52 \cdot 10^{-2}$, close to the value $\epsilon \approx 1.16\cdot 10^{-2}$ of the global scenario. Lastly, for $\alpha = 1$, we achieve a very low error rate $\epsilon<10^{-4}$, not shown in the figure.

\section{Conclusions and Future Work}

In this paper, we investigated different solutions to exploit the correlation among \ac{uwac} features to authenticate underwater network nodes.
Our results suggest that it is advantageous to share some additional information on local channel features with the sink node for a more accurate final decision rather than simply merging local decisions.
{Future work includes, among others: a) the development of \ac{ml} models able to authenticate the users in time-varying channels that can be trained using channel features from legitimate transmitters, and b) more sophisticated attacks and defense strategies, also based on generative adversarial networks.}

\IEEEtriggercmd{\enlargethispage{-15cm}}
\IEEEtriggeratref{2}

\bibliographystyle{IEEEtran}
\bibliography{IEEEabrv,biblio}    

\end{document}